\documentclass{emulateapj}

\newcommand{\fign}[3]
{
        \begin{figure}
        \includegraphics[width=3.25in]{#1}
        \caption{#2}
        \label{#3}
        \end{figure}
}

\newcommand{\figw}[3]
{
        \begin{figure*}
        \begin{centering}
        \includegraphics[width=5in]{#1}
        \caption{#2}
        \label{#3}
        \end{centering}
        \end{figure*}
}

\begin{document}
\title{Variability Profiles of Millisecond X-Ray Pulsars: Results of Pseudo-Newtonian 3D MHD Simulations}
\author{A.K. Kulkarni}
\affil{Department of Astronomy, Cornell University, Ithaca, NY-14853; akshay@astro.cornell.edu}
\author{M.M. Romanova}
\affil{Department of Astronomy, Cornell University, Ithaca, NY-14853-6801; romanova@astro.cornell.edu}

\begin{abstract}
We model the variability profiles of millisecond period X-ray
pulsars. We performed three-dimensional magnetohydrodynamic
simulations of disk accretion to millisecond period neutron stars
with a misaligned magnetic dipole moment, using the pseudo-Newtonian Paczy\'nski-Wiita potential to model general relativistic effects. We found that the shapes
of the resulting funnel streams of accreting matter and the hot
spots on the surface of the star are quite similar to those for
more slowly rotating stars obtained from earlier simulations using the Newtonian potential. The funnel
streams and hot spots rotate approximately with the same angular
velocity as the star. The spots are bow-shaped (bar-shaped) for
small (large) misalignment angles. We found that the matter falling on the star has a higher Mach number when we use the Paczy\'nski-Wiita potential
than in the Newtonian case.

Having obtained the surface distribution of the emitted flux, we
calculated the variability curves of the star, taking into account
general relativistic, Doppler and light-travel-time effects. We
found that general relativistic effects decrease the pulse
fraction (flatten the light curve), while Doppler and
light-travel-time effects increase it and distort the light curve. We also found that the light curves from our hot spots are reproduced reasonably well by spots with a gaussian flux distribution centered at the magnetic poles.
We also calculated the observed image of the star in a few cases,
and saw that for certain orientations, both the antipodal hot
spots are simultaneously visible, as noted by earlier authors.
\end{abstract}


\section{Introduction}

Millisecond X-ray pulsars show bursts of periodic and
quasi-periodic variability in the X-ray (Stella \& Vietri 1999;
van der Klis 2000; Chakrabarty et al. 2003; Wijnands et al. 2003).
Six of these pulsars are believed to be accretion-powered
(Wijnands 2005). Numerical modelling of
accretion is an important tool for studying these phenomena.
Three-dimensional magnetohydrodynamic (3D MHD) simulations of disk
accretion to rotating magnetized stars with a misaligned dipole
magnetic field have been carried out by Koldoba et al. (2002) and
Romanova et al. (2003; 2004, hereafter - R04). They found that the
accreting matter is channelled by the star's magnetic field into
two antipodal streams or funnels, and falls on the stellar surface
forming two antipodal hot spots -- regions of relatively high
temperature. Such flows have been predicted theoretically (e.g.,
Pringle \& Rees 1972; Lamb, Pethick \& Pines 1973; Ghosh \& Lamb
1978, 1979), but were modelled numerically only recently. Assuming
that the energy of the infalling matter is converted entirely into
radiation, R04 then calculated the variability curves of the star.
Those simulations and calculations were performed for a generic
star in a completely classical framework. We refined those
simulations, focusing on rapidly rotating ($\sim$ 3-5 ms period)
neutron stars. For such stars, the following issues arise: (1) For
compact objects like neutron stars, general relativistic effects
significantly influence the accretion process as well as the
observed flux. (i) We modelled general relativistic effects on
accretion by using the pseudo-Newtonian Paczy\'nski-Wiita
potential (Paczy\'nski and Wiita 1980) which reproduces some
important features of the Schwarzschild geometry, like the
positions of the innermost stable and marginally bound circular
orbits. (ii) The variability curve of the star also changes
significantly, because gravitational bending of light emitted by
the star allows more of the star than the hemisphere facing the
observer to be visible, and gravitational redshift of the light
decreases the total flux observed. We use the Schwarzschild metric
to take these effects into account. (2) For rapidly rotating stars
like millisecond pulsars, the rotation of the star changes the
observed flux through the twin special relativistic effects of
Doppler shift and relativistic beaming of the emitted radiation.
Henceforth we refer to these two effects collectively as the
``Doppler effect.'' (3) The time difference between light emitted
from different points on the star reaching the observer is
important when the linear speed of the emitting region is
comparable to the speed of light, and also when the emitting
object is compact, and causes distortion of the observed shape of
the hot spots (the apparent position of a point on the stellar
surface can differ by as much as $10^\circ$ from its actual
position). We build upon the earlier calculations to take these
effects into account. Additionally, due to the high rotation
speed, the Kerr metric would be closer to the actual metric around
the star than the Schwarzschild metric. However we do not use the
Kerr metric, since we find from numerical integrations that the
frame dragging effects introduced by the Kerr metric are
relatively small (see also Braje, Romani \& Rauch 2000),
particularly when compared with the other errors introduced by the
assumptions in our variability model.

These effects on the light curves have been taken into account by
earlier authors to obtain light curves for simple hot spots (see,
e.g., Pechenick, Ftaclas \& Cohen 1983; Ftaclas, Kearney \&
Pechenick 1986; Braje, Romani \& Rauch 2000;
Ford 2000; Beloborodov 2002; Poutanen \& Gierlinski 2003;
Viironen \& Poutanen 2004; Bhattacharyya et al. 2005). Here, we
obtain light curves using realistic hot spots obtained from our
accretion simulations.

In section 2 we present the results of our MHD simulations. We
then discuss the analytical background for calculating variability
curves in section 3, and show the variability curves for synthetic
and realistic spots in sections 4 and 5 respectively, followed by
some concluding remarks in section 6.

\section{Disk Accretion - 3D Simulations}
\subsection{Earlier Simulations}
We briefly describe earlier 3D MHD simulations (Koldoba et al.
2002; Romanova et al. 2003; R04). The star has a dipole magnetic
field, the axis of which makes an angle $\Theta$ with the star's
rotation axis. The rotation axes of the star and the accretion
disk are aligned. The disk has a low-density corona which also
rotates about the same axis. To model stationary accretion, the
disk was chosen to initially be in a quasi-equilibrium state,
where the gravitational, centrifugal and pressure gradient forces
are in balance (Romanova et al., 2002). Viscosity is modelled
using the $\alpha$-model (Novikov \& Thorne 1973; Shakura \&
Sunyaev 1973). To model accretion, the ideal MHD equations were
solved numerically in three dimensions, using a Godunov-type
numerical code, written in a ``cubed-sphere'' coordinate system
rotating with the star (Koldoba et al., 2002; Romanova et al.,
2003). The boundary conditions at the star's surface amount to
assuming that the infalling matter passes through the surface of
the star. So the dynamics of the matter after it falls on the star
was ignored. It was found that the inward motion of the accretion
disk is stopped by the star's magnetosphere at the Alfv\'en
radius, where the magnetic and matter energy densities become
equal. At that point the matter leaves the disk and moves along
the magnetic field lines. This flow is called a funnel stream. In
this region, the matter radiates primarily in the X-ray. It heats
up the star's surface where it falls, forming ``hot spots.'' There
are two antipodal funnel streams and hot spots. From the point of view of an external observer, they rotate with
approximately the same angular velocity as the star, causing the
observed flux to vary periodically with time. The shape of the funnel streams and hot spots keeps changing slightly with time, leading to quasi-variability in the observed flux.

\subsection{Reference Values}
In our new simulations, we use the same model as described above.
The simulations are done using the following dimensionless
variables: the radial coordinate $r' = r/R_0$, the fluid velocity
${\mathbf v}' = {\mathbf v}/v_0$, the density $\rho' =
\rho/\rho_0$, the magnetic field ${\mathbf B}' = {\mathbf B}/B_0$,
the pressure $p' = p/p_0$, the temperature $T' = T/T_0$, and the
time $t' = t/t_0$. The variables with subscript 0 are dimensional
reference values and the unprimed variables are the dimensional
variables. Because of the use of dimensionless variables, the
results are applicable to a wide range of objects and physical conditions,
each with its own set of reference values. To apply our simulation results to a particular situation, we have the freedom to choose three parameters, and all the reference values are calculated from those. We choose the mass, radius and surface magnetic field of the star as the three independent parameters.

The reference values are determined as follows: The unit of
distance $R_0$ is chosen such that the star has radius $R =
0.35R_0$. The reference velocity is the Keplerian velocity at
$R_0$, $v_0 = (GM/R_0)^{1/2}$, and $\omega_0 = v_0/R_0$ is the
reference angular velocity. The reference time is $t_0 = R_0/v_0$.
The reference surface magnetic field of the star is $B_{\star_0}$.
The reference magnetic field, $B_0$, is the initial magnetic field
strength at $r=R_0$, assuming a surface magnetic field of
$B_{\star_0}$. The reference density is taken to be $\rho_0 =
B_0^2/v_0^2$. The reference pressure is $p_0 = \rho_0 v_0^2$. The
reference temperature is $T_0 = p_0/{\mathcal R}\rho_0$, where
$\mathcal{R}$ is the gas constant. The reference accretion rate is
$\dot{M}_0 = \rho_0 v_0 R_0^2$. The reference energy flux is
$\dot{E}_0 = \rho_0 v_0^3R_0^2$. The reference value for the
effective blackbody temperature of the hot spots is $(T_{\mathrm
eff})_0 = (\rho_0 v_0^3/\sigma)^{1/4}$, where $\sigma$ is the
Stefan-Boltzmann constant.

For the millisecond pulsars in our simulations, we take the mass
of the neutron star to be $M = 1.4 M_\odot = 2.8 \times 10^{33}
\mbox{ g}$ and its radius $R = 10 \mbox{ km} = 10^6 \mbox{ cm}$.
The reference length scale is $R_0 \approx 2.86 R = 2.86 \times
10^6 \mbox{ cm}$. The reference velocity is $v_0 = 8.1 \times 10^9
\mbox{ cm s}^{-1}$. The reference time is $t_0 = 0.35 \mbox{ ms}$.
The reference surface magnetic field is $B_{\star_0} = 10^8 \mbox{
G}$, which is a typical value for millisecond pulsars. Then the
reference magnetic field is $B_0 = B_{\star_0}(R/R_0)^3 \approx
4.3 \times 10^6 \mbox{ G}$. The reference density is $\rho_0 = 2.8
\times 10^{-7} \mbox{ g cm}^{-3}$. The reference pressure is $p_0
= 1.8 \times 10^{13} \mbox{ dynes cm}^{-2}$. The reference
temperature is $T_0 = 7.9 \times 10^{11} \mbox{ K}$. The reference
value of the effective blackbody temperature is $(T_{\mathrm
eff})_0 \approx 7.2 \times 10^6 \mbox{ K}$. The reference mass
accretion rate is $\dot{M}_0 \approx 1.85 \times 10^{16} \mbox{ g
s}^{-1} \approx 2.9 \times 10^{-10} M_\odot \mbox{ yr}^{-1}$. The
reference energy flux is $\dot{E}_0 \approx 1.2 \times 10^{36}
\mbox{ erg s}^{-1}$.

Subsequently, we drop the primes on the dimensionless variables
and show dimensionless values in the figures.

\figw{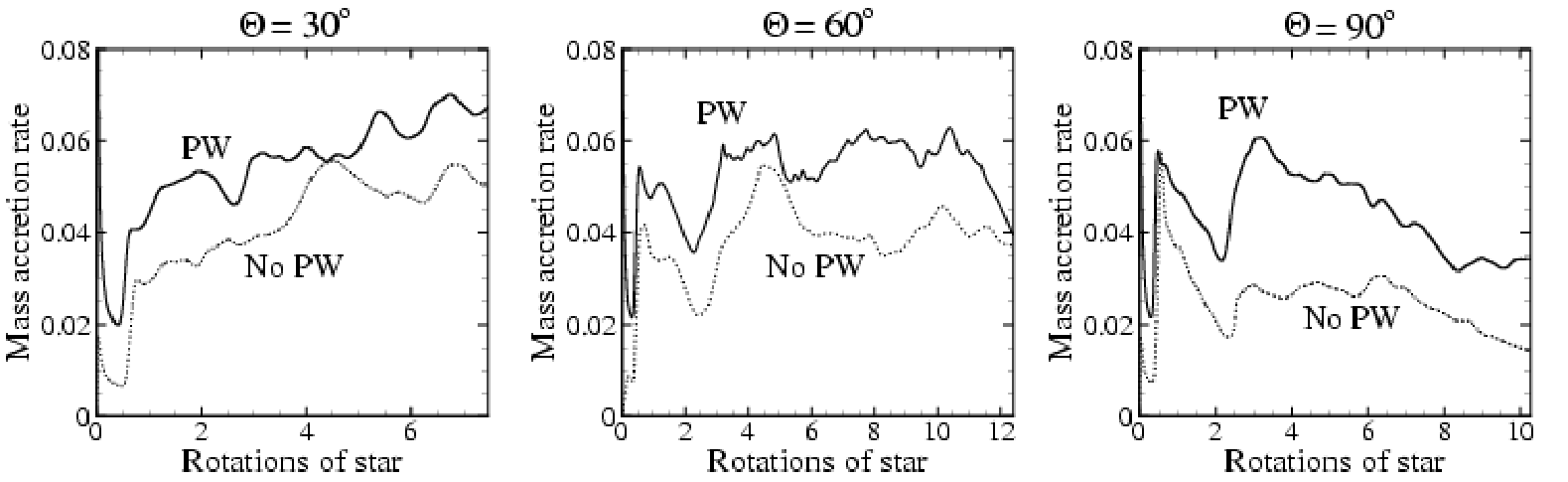}{Mass accretion rates for a 3-ms pulsar with (solid lines) and without (dotted lines) the Paczy\'nski-Wiita potential, for different misalignment angles.}{pm}

\figw{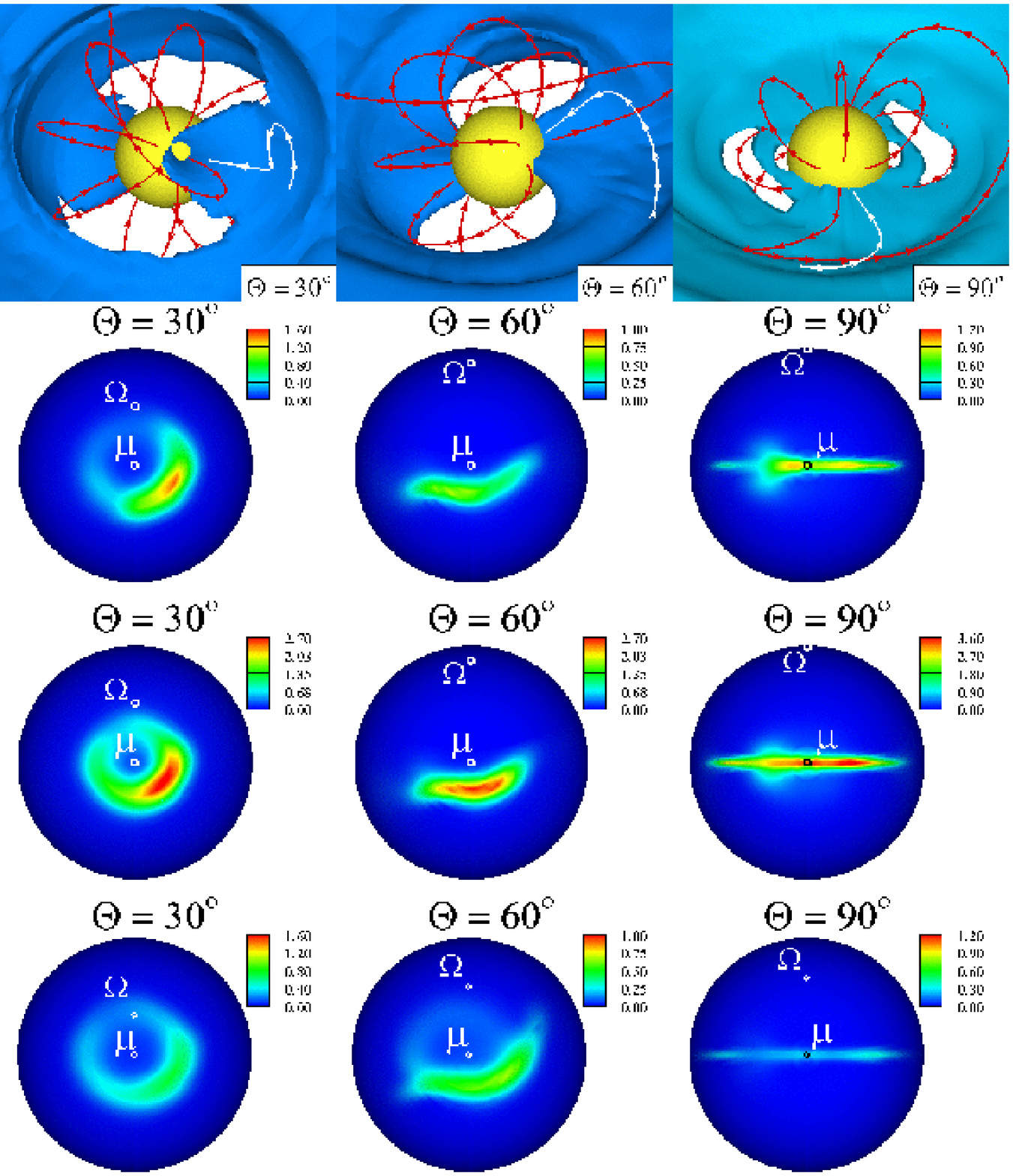}{{\it Top row:} Matter flow around a star with $P=3\mbox{ms}$ in the ``PW'' case for different misalignment angles. The red lines are sample magnetic field lines and the white lines are sample streamlines of matter flow. {\it Second row:} Distribution of emitted flux on the star's surface, without using the PW potential. {\it Third row:} Distribution of emitted flux on the star's surface, using the PW potential. {\it Bottom row:} Distribution of emitted flux on a sphere of radius 15 km, using the PW potential.}{funnelspots}

\subsection{New Simulations}

We performed new simulations for rapidly rotating (3-5 ms period)
neutron stars. We followed the same procedure as described above,
with two major changes: (1) We modelled general relativistic
effects on accretion by using the pseudo-Newtonian
Paczy\'nski-Wiita (PW) potential (Paczy\'nski \& Wiita 1980),
$\Phi(r) = -GM/(r-r_g)$, where $M$ is the mass of the star and
$r_g = 2GM/c^2$ is its Schwarzschild radius. (2) We calculated the
variability curves taking into account relativistic and
light-travel-time effects.

We used the following parameters in our simulations: The surface
magnetic field of the star was $5 \times 10^7$ G. The star's
Schwarzschild radius was $r_g = 4.15 \mbox{ km } = 4.15 \times
10^5 \mbox{ cm}$, or 0.145 in dimensionless units. The disk and corona initially had temperatures
of 0.01 and 1 respectively, and densities of 1 and 0.01
respectively. The disk was thinner than in the earlier
simulations. The viscosity $\alpha$-parameter was 0.04. Each of
the six blocks of our cubed-sphere grid had 65 cells in the radial
direction and 31 in the angular direction, which corresponds to an
outer disk radius of $\sim 8.7$, or about 25 stellar radii.

We used a modified code and
modified initial conditions that use the PW potential instead of the
Newtonian one. The initial values of physical variables in the
disk do not change significantly due to use of the PW potential.
We found that the shapes of the funnel streams and hot spots are
similar to those in earlier simulations for more slowly rotating
stars, and in simulations with the Newtonian potential. We also
found that the velocity of the infalling matter near the surface
of the star is higher in the ``PW'' case than in the ``non-PW''
case. This is expected because the PW potential is stronger
than the Newtonian one.

The accretion rate is higher in the PW case than in the
non-PW case, as Fig. \ref{pm} shows. This is again expected because the PW potential is stronger. This will lead to faster depletion of the inner disk matter. If angular momentum is efficiently transported outward, e.g., by the magneto-rotational instability (see, e.g., Hawley \& Krolik 2001), then the influence of the PW potential will eventually be felt in the outer disk regions, and enhanced accretion could be sustained for a longer time.

The top row of Fig. \ref{funnelspots} shows the flow of matter
around a star with $P=3\mbox{ ms}$ in the PW case. The flow is
similar in the non-PW case. The next two rows show the hot
spots formed on the surface of the star, without and with the PW potential. We see that the
position and shape of the hot spots depends on the misalignment
angle, but does not significantly depend on the presence of the PW
potential. The hot spots are bow shaped (bar-shaped) for small
(large) misalignment angles. The emitted flux is highest at the
center of the spots, and decreases outwards. Note that the hot spots are not usually centered at the magnetic poles. In fact, they do not even fall on the magnetic poles in most cases. We see that the emitted flux is higher in the PW case, which is expected because the PW potential is stronger than the Newtonian one.

To get an idea of the conditions at the surface of a larger neutron star with the same mass, rotation period and magnetic dipole moment, in a similar situation, we can look at the surface of a sphere of radius 15 km concentric with the star in our simulations. This approach is valid because changing the star's radius does not change the accretion flow around the star, since the PW potential depends only on the star's Schwarzschild radius. The bottom row of Fig. \ref{funnelspots} shows the hot spots on such a sphere. We see that, other conditions remaining the same, a larger star has much fainter hot spots, which is again to be expected because the accreting matter has a lower velocity at the surface of the larger star.

For 3-ms pulsars, the typical values of physical quantities
observed in our simulations after 4-6 rotations of the star are as
follows: The surface magnetic field does not change in our model,
and hence is $\sim 5 \times 10^7$ G. The mass accretion rate to
the star $\sim 10^{15}$ g s$^{-1} \sim 10^{-12}M_\odot$ yr$^{-1}$.
The total power emitted from the star, after correcting for
gravitational redshift, $\sim 10^{34}$ erg s$^{-1}$. In the hot
spots, the matter density $\sim 10^{-7}$ g cm$^{-3}$. The speed of
the inflowing matter $\sim$ 2 $\times 10^{10}$ cm s$^{-1}$. The
effective blackbody temperature of the hot spots $\sim 5 \times
10^6$ K. The matter pressure $\sim 10^{12}$ dynes cm$^{-2}$. The
Mach number $\sim 1-6$ in the PW case and $\sim 0.8-3$ in the
non-PW case. The surface distributions of the density, pressure,
velocity and temperature closely follow that of the emitted
flux.

\section{Calculation of the variability curves}

Having obtained the distribution of emitted flux on the star's
surface, we can calculate the flux received by an observer.
Without including the relativistic and light-travel-time effects,
the observed flux at a large distance $D$ from the star is
proportional to
\begin{equation}
\label{noeff}
J = \int_{\cos \psi > 0} dS\,\,I_E(\mathbf R, \psi) \cos \psi,
\end{equation}
where $dS$ is an element of area of the star's surface at a position $\mathbf R$, and $I_E(\mathbf R, \psi)$ is the intensity of radiation emitted by
$dS$ at an angle $\psi$ with respect to the local radial direction (Fig. \ref{lighttravel}a). We have ignored the overall
factor of $1/D^2$ in the flux.
When calculating this numerically, the
integral becomes a sum over grid elements. We now consider the
gravitational and rotational effects one by one. These calculations have been done by earlier authors (see, e.g., Poutanen \& Gierlinski 2003; Viironen \& Poutanen 2004), but we present them here for the sake of completeness and for discussing slight differences in approach.

\fign{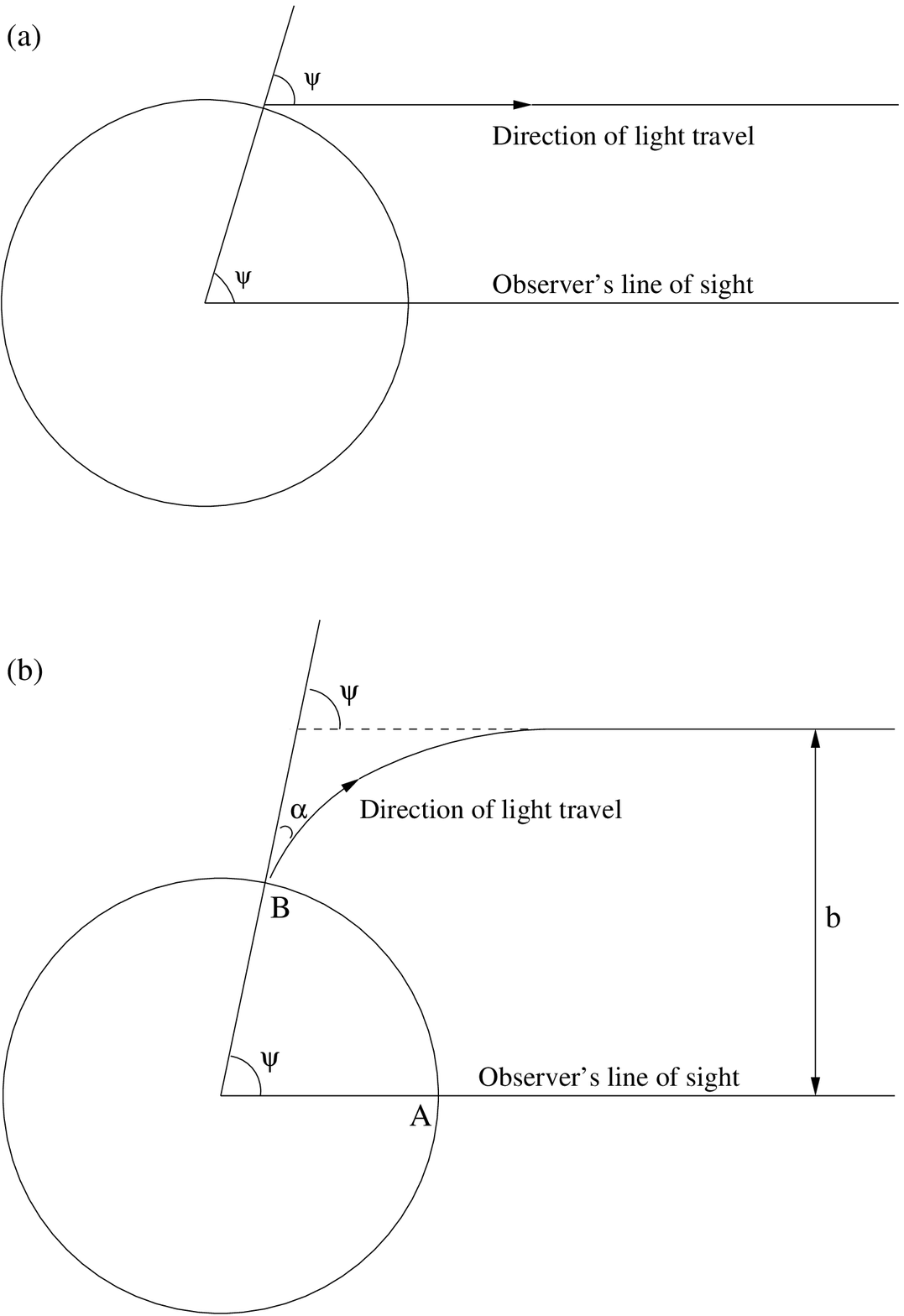}{Path of light emitted by a compact object: (a) neglecting light bending, and (b) including light bending.}{lighttravel}

\subsection{General Relativistic Effects}
When taking general relativistic effects into account, two
approaches are possible towards calculating the observed flux. In
the traditional ray tracing method (see, e.g., Braje, Romani \&
Rauch 2000; Bhattacharyya et al. 2005), one traces light rays
backwards from the observer's image plane to the star's surface,
by numerically integrating the geodesic equations. The flux
emitted from the point where a light ray meets the star will then
determine the intensity of that ray. This can be called an
``observer-centered'' approach. For our purposes, however, it is
more convenient to take a ``grid-centered'' approach, since our
MHD simulations give us the hot spot data on a grid, and it is
more convenient to calculate the contribution of each grid element
to the observed flux, and then sum over all grid elements, as
stated above. So we use the following approach (see, e.g.,
Beloborodov 2002).

Because of light bending, the light from point B in Fig.
\ref{lighttravel}b needs to be emitted at an angle $\alpha$ such
that after bending, it travels towards the observer, that is, it
travels at an angle $\psi$ with respect to the original radial
direction. To calculate the observed flux we need a relation
between $\psi$ and $\alpha$. We use the Schwarzschild metric. From
the geodesic equations, we then have (see, e.g., Misner, Thorne \&
Wheeler 1973; Pechenick, Ftaclas \& Cohen 1983)
\begin{equation}
\label{psialpha} \psi = \int_R^\infty
\frac{dr}{r^2}\left[\frac{1}{b^2} - \frac{1}{r^2}
\left(1-\frac{r_g}{r}\right)\right]^{-1/2},
\end{equation}
where $r$ is the radial Schwarzschild coordinate, $R$ and $r_g$
are the star's radius and Schwarzschild radius respectively, and
$b$ is the impact parameter. Using the four-velocity of a photon
at the point of emission, we can relate $b$ to $\alpha$ to get
(see, e.g., Beloborodov 2002) $b=R(1-r_g/R)^{-1/2}\sin\alpha.$ We
now have the desired relation between $\psi$ and $\alpha$. The
integration in equation (\ref{psialpha}) cannot be done
analytically, but is relatively easy to do numerically. However,
the problem in our case is a more difficult one, because in the
grid-centered approach we know $\psi$ for each grid element, and
we have to solve equation (\ref{psialpha}) for $\alpha$.
Computationally, this is very difficult to do exactly. So we use
the cosine relation, an approximate relation (due to Beloborodov
2002), which has the advantages of being very simple to use and
highly accurate: 
\begin{equation}
\label{bending}
1 - \cos \alpha \approx (1 - \cos \psi)\left(1-\frac{r_g}{R}\right).
\end{equation}
In that case the observed flux is given by (Beloborodov 2002)

\begin{equation}
\label{fluxold} J = \left( 1-\frac{r_g}{R} \right)^2
\int_{\cos \alpha > 0} dS\,\,I_E(\mathbf R, \alpha) \cos \alpha.
\end{equation}
So the angle $\psi$ in equation (\ref{noeff}) is replaced by
$\alpha$, and we get a prefactor of $(1-r_g/R)^2$ due to
gravitational redshift.

An interesting consequence of light bending is that the observer
can see some radiation from the far side of the star (see, e.g.,
Beloborodov 2002). In particular, both the antipodal hot spots of
pulsars can be seen simultaneously in some cases.

\subsection{Doppler Effect}
The above discussion is valid if $I_E(\mathbf R, \alpha)$ is the intensity
in a reference frame which is at rest with respect to the
observer. However, our MHD simulations calculate the intensity in a reference frame which is rotating with the
star. We need to relate the intensities in these two frames. We use the
invariance of $I_{\nu}/\nu^3$ along a ray of light, where
$I_{\nu}$ is the specific intensity, or the intensity per unit
frequency range. Then for a ray emitted from a point on the star
which is moving with a velocity $v=\beta c$ and Lorentz factor
$\gamma=(1-\beta^2)^{-1/2}$, we have\footnote{Notation in this
subsection: Unprimed quantities are those measured in a frame at
rest (with respect to the observer) at the star's surface. Primed
quantities are those measured in a frame rotating with the star.}
$I_{\nu} = I_{\nu'}'/\gamma^3(1-\beta\mu)^3$, since $\nu$
and $\nu'$ are related by the Doppler formula $\nu' =
\nu\gamma(1-\beta\mu)$. Here $\mu=\cos \theta$, where $\theta$ is
the angle, as measured in the unprimed frame, between the
direction of emission of the light and the direction of motion of
the emitting surface element of the star. The direction of
emission of light is given by equation (\ref{bending}).
Integrating over frequency, we then have $I = I'/\gamma^4(1-\beta\mu)^4$. Thus in equation
(\ref{fluxold}) for the observed flux, we pick up an extra factor
of $1/\gamma^4(1-\beta\mu)^4$. Also, the areas of the hot spots as
measured by photon beams in the two frames are related by $dS' =
\gamma(1-\beta\mu)dS$ (Terrell 1959). This is a consequence of the projected area $dS \cos \alpha$ being a Lorentz invariant. So the flux is now given by

\begin{equation}
\label{fluxnew} J = \left( 1-\frac{r_g}{R} \right)^2
\int_{\cos \alpha > 0} dS'\frac{1}{\gamma^5(1-\beta\mu)^5}
I_E'(\mathbf R', \alpha') \cos \alpha.
\end{equation}

\subsection{Light Travel Time Effects}
Now we shall take into account the fact that light from different
parts of the star takes different amounts of time to reach the
observer. We need to use the general relativistic expression for
the light travel time, since we are dealing with a compact object.
We again use the Schwarzschild metric. Using the geodesic
equations, the time difference between light emitted from points A and B in
Fig. \ref{lighttravel}b reaching the observer is (assuming $D
\gg R$)
\begin{equation}
\label{deltat} \delta t(b) = \frac{1}{c} \int_R^\infty \left.
\frac{dr}{\left(1-\frac{r_g}{r}\right)} \left\{
\left[1-\frac{b^2}{r^2}\left(1-\frac{r_g}{r}\right)\right]^{-1/2}
- 1 \right\} \right. .
\end{equation}
We then need to find the apparent position of each grid element at
a given time. At the time when the point A in Fig. \ref
{lighttravel}b will appear to be at the center of the observer's
image, the point B will appear to be not where it is shown in the
figure, but where it was a time $\Delta t$ ago, where $\Delta t$
must satisfy
\begin{equation}
\label{deltatimp} \delta t(b_{\Delta t}) = \Delta t .
\end{equation}
Here $b_{\Delta t}$ is the impact parameter for light emitted from
point B a time $\Delta t$ ago. This determines the apparent
position of grid element B at the time when grid element A is at
the position shown in Fig. \ref{lighttravel}b.

In the observer-centered approach, one needs to solve equation
(\ref{deltat}), which is relatively easy to do numerically.
However in the grid-centered approach, one has to solve equation
(\ref{deltatimp}) which, after substituting for $\delta
t(b_{\Delta t})$ from equation (\ref{deltat}), becomes an integral
equation which cannot be solved analytically, and is very
difficult to solve numerically. So, following Beloborodov(2002),
we expand the integrand in equation (\ref{deltat}) in powers of $x
= (1-\cos \psi) = (1-\cos \alpha)/(1-r_g/R)$ and then perform the
integration. We keep as many terms as are needed for sufficient
accuracy (accuracy can be checked by comparing the values obtained
from the series with those obtained by integrating eq.
(\ref{deltat}) numerically). The resulting equation, although
still implicit in $\Delta t$, is much easier to solve numerically.
We can now calculate the apparent position of any grid point at
any time, and then compute the flux as given by equation
(\ref{fluxnew}) by summing over all grid elements.

\subsection{Frame Dragging Effects}
The neutron stars considered in our simulations have $\mbox{R}=10
\mbox{ km}$ and $\mbox{M}=1.4\mbox{M}_\odot$. The fastest rotators
we considered have $\mbox{P}=3 \mbox{ ms}$. For these parameters,
we tried to find out the significance of frame dragging by
numerically evaluating frame dragging corrections to the path of
light in the equatorial plane of the star, using the Kerr metric. We found that
corrections to the angle $\alpha$ (Fig. \ref{lighttravel}b) are at
most $\sim 4^\circ$. Corrections to the time delay $\delta t$ (eq.
\ref{deltat}) are of the order of a few percent, which is not
large considering the fact that the effect of time delay on the
variability curve is itself quite small, as we shall see in
section 5. The gravitational redshift factor has corrections of
$\lesssim 1\%$. We thus expect the errors introduced by ignoring
frame dragging to be much smaller than those introduced by the assumptions in our variability model (which we discuss in section 5). This, together
with the fact that frame dragging effects are very difficult to
take into account numerically, is why we do not do so. Variability
curves have been calculated by earlier authors taking these
effects into account (Braje, Romani \& Rauch 2000), and they found
that the curves do not change significantly due to these effects.

\section{Examples with simple synthetic spots}

\subsection{Point Spot}
It is clear from the foregoing discussion that general
relativistic effects become stronger with increasing $r_g$, and
Doppler and time delay effects become stronger with increasing
rotation speed of the star. To see clearly the effect of all these
effects on the light curve, it is useful to consider the simple
case of a point spot shown in Fig. \ref{pointspot}. We have only
one spot, which is on the rotational equator of the star, and the
observer is in the equatorial plane of the star. When no effects
are included, the light curve is simply a half-sinusoid (due to
the $\cos \psi$ factor in eq. \ref{noeff}) with a maximum when the
spot is at point C, and minima when the spot is on the far side of
the star (Fig. \ref{timedelay}). The points labeling the curve
correspond to the spot being at the respective points in Fig.
\ref{pointspot}. Let us now look at all the effects separately.

{\bf Time delay.} Referring to Fig. \ref{pointspot}, the light
from, say, point B, takes longer to reach the observer than that
from point C. So in the light curve, B shifts to B', and so on.
Due to this, the light curve gets distorted as shown. The peak
position of the time-delay curve is arbitrary, because the choice
of the reference point for the time delay (point C in this case)
is arbitrary. So time delay ``bends peaks to the left''. Note that
in any case, no shift in peak position is observable, since the
observer sees only one light curve, the one including all effects.
We discuss the shifts here merely to understand our results the
better.

\fign{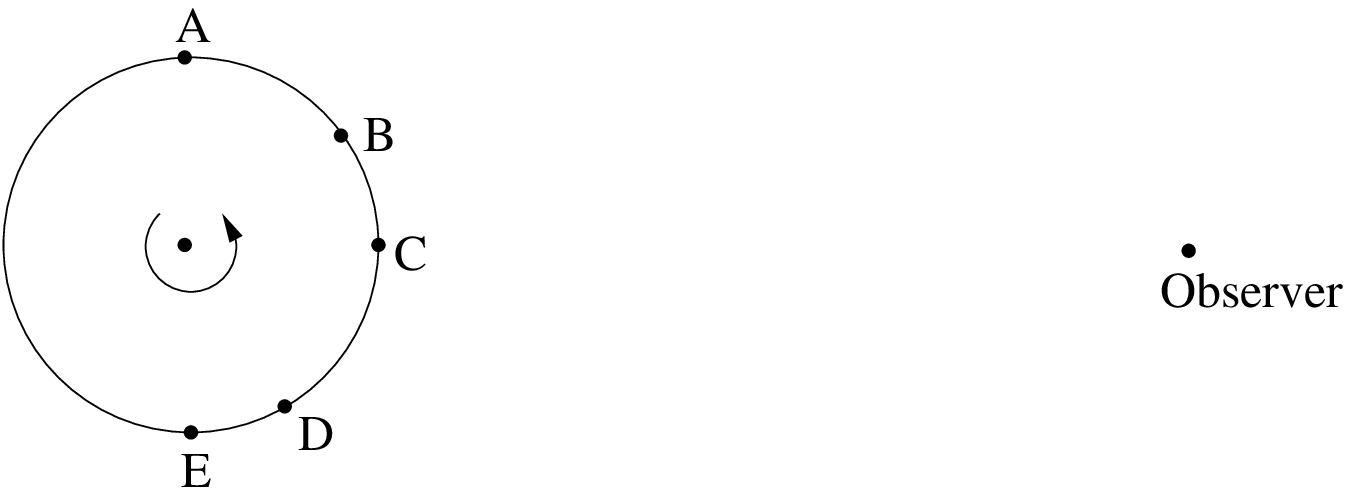}{Geometry for the case of a single point spot on the equator with the observer in the equatorial plane.}{pointspot}

\fign{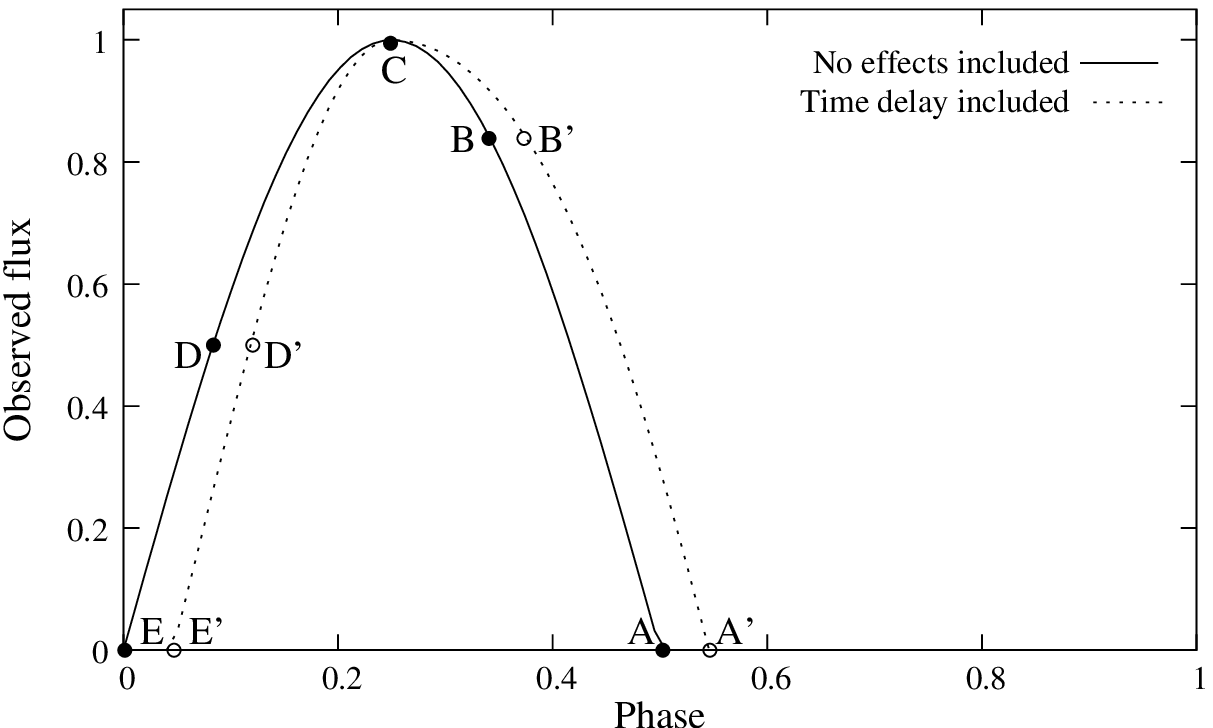}{Schematic light curve for the case shown in Fig. \ref{pointspot} with and without including time delay, over a period of 1 rotation of the star. The flux has arbitrary units.}{timedelay}

{\bf Doppler.} In the above case, the observed flux from a certain
point is determined only by the $\cos \psi$ factor in equation
(\ref{noeff}).When we include the Doppler effect, relativistic
beaming comes into the picture. In our simple case, beaming of
radiation towards the observer is maximum at point E, while $\cos
\psi$ is maximum at point C. So the peak occurs at some
intermediate point D. The Doppler effect thus shifts the peak of
the light curve. The intensity of the peak also increases
dramatically, because the spot has a significant velocity along
the observer's line of sight at that point. For other geometries
where the spot always moves almost perpendicular to the line of
sight, beaming reduces the peak intensity.

{\bf General relativity.}  Without general relativistic effects,
the visible portion of the star is determined by $\cos \psi > 0$.
When we include general relativistic effects, this condition
changes to $\cos \alpha >0$, with $\alpha$ given by equation
(\ref{bending}). Since $\cos \alpha > \cos \psi$, this condition
implies that more than half of the stellar surface is visible to
the observer at any time. For certain geometries and with
antipodal hot spots, this makes both the spots visible
simultaneously. In such a case, if the spots are identical, the
observed flux is almost constant for the duration for which both
spots are visible (see Beloborodov (2002) for a detailed
discussion). Even when only one spot is visible, the reduced
modulation of the flux (due to $\cos \alpha > \cos \psi$)
flattens the light curve. The peak position does not change,
however, since the peak occurs when $\cos \alpha$ is maximum,
which is when $\cos \psi$ is maximum. Gravitational redshift
decreases the total observed flux, except when both the antipodal
spots are simultaneously visible, in which case the observed flux
can increase.

\subsection{Gaussian Spots}

\fign{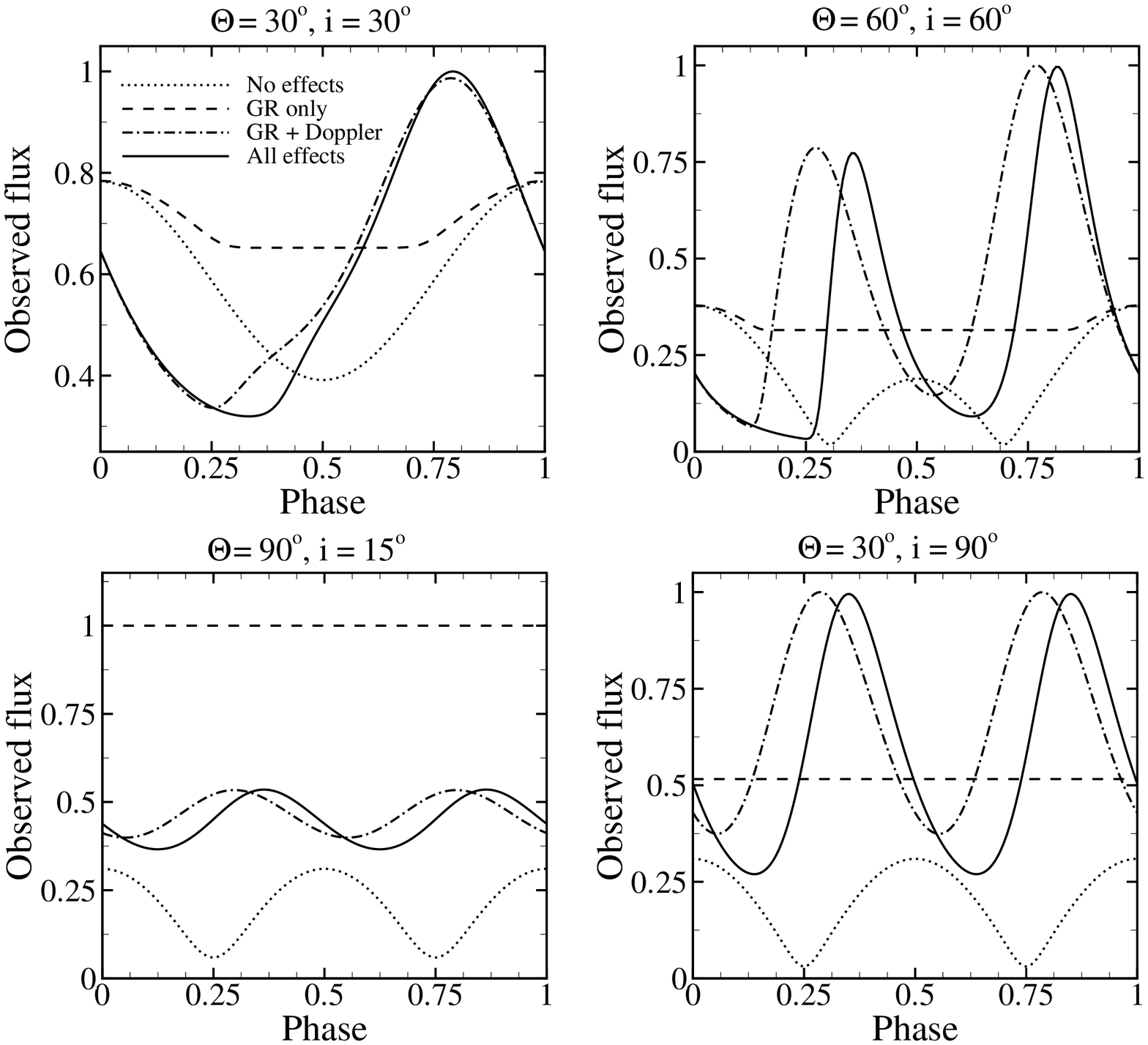}{Examples of relativistic and light-travel-time effects on the light curves for gaussian spots. Dotted curves do not include any effects, dashed curves include only GR effects, dash-dotted curves include GR and Doppler effects, and solid curves include all effects. The fluxes are normalized.}{gaussian}

To illustrate the above effects, we show some variability curves
for spots with a gaussian flux distribution, maximum at the center
and tapering outwards. We consider a hypothetical neutron star
with $R=10\mbox{ km}$ and $M=1.4M_\odot$. To bring out clearly the effects of rotation, we choose a very small rotation period $P=0.5\mbox{ ms}$. The
spots have a width of $10^\circ$. Fig. \ref{gaussian} shows some
variability curves for different misalignment angles $\Theta$ and
observer inclination angles $i$. (The inclination angle is the
angle between the observer's direction and the star's rotation
axis.) We show the curves over one rotation period of the star
since, because of the assumptions in our variability model, the
curves are perfectly periodic with a period equal to the star's
spin period. Without including any effects, the light curve has
one or two peaks depending on whether one or both the hot spots
are seen during one rotation period (see R04 for a detailed discussion). General relativistic effects
reduce the pulse fraction and, in the cases where both hot spots
are simultaneously visible for some time, completely flatten the
light curve for that duration. The time delay effect is seen to
distort the light curves. We also see that the Doppler and time
delay effects increase the pulse fraction. These two effects are
strongest when the hot spots are moving almost directly towards or
away from the observer, which happens at large inclination angles.

\section{Variability curves for realistic spots}

\fign{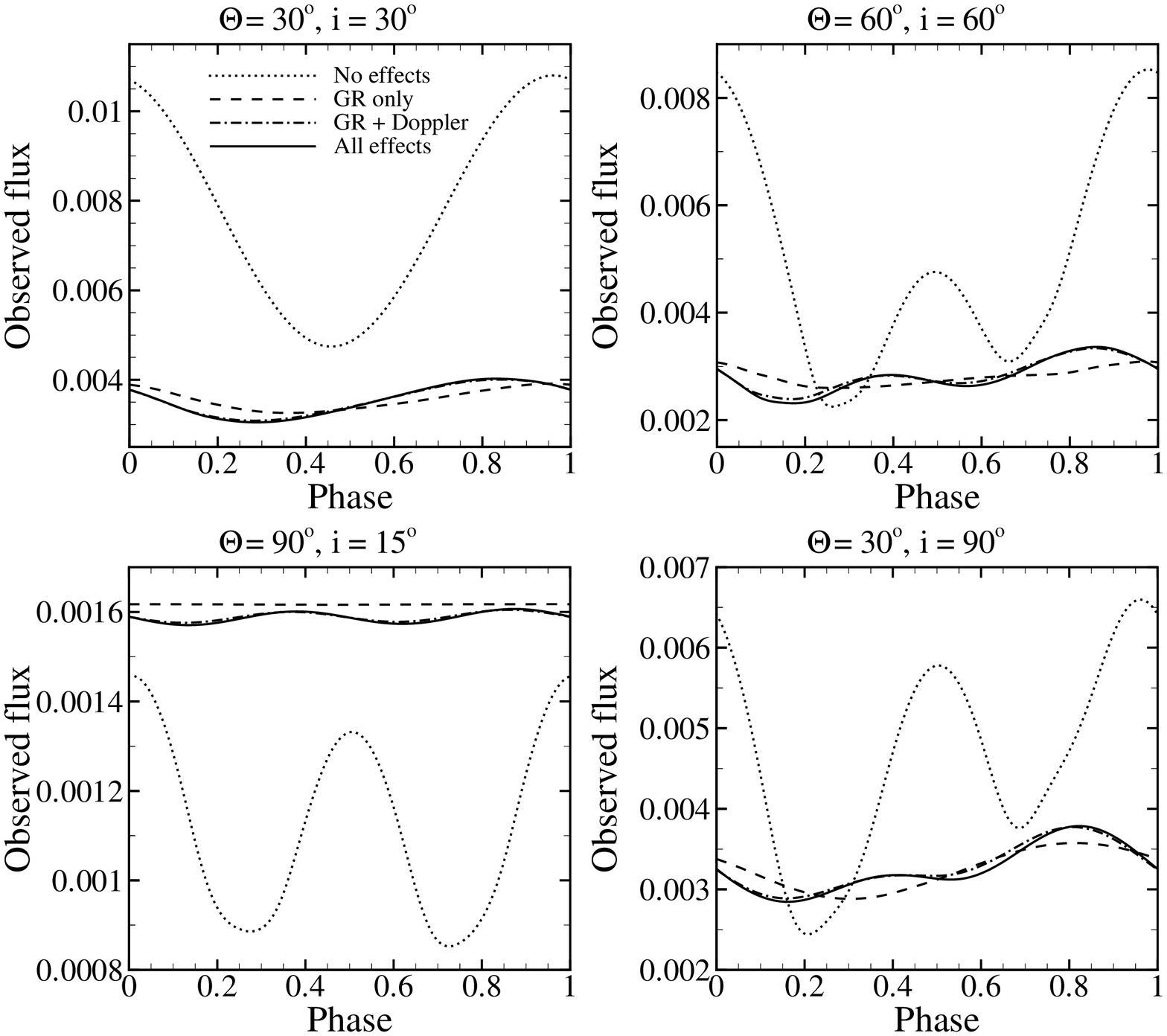}{Examples of relativistic and light-travel-time effects on the light curves of 3-ms pulsars for realistic spots in the PW case. The line patterns mean the same as in Fig. \ref{gaussian}.}{real}

We now turn our attention to the realistic spots discussed in section 2. To calculate the variability curves, we make the following assumptions: only
radiation from the hot spots contributes to the light curve; when
the infalling matter falls on the star, its entire kinetic and
thermal energy is converted into radiation, which is emitted
isotropically; the emitted radiation does not interact with the
accreting matter. Also, recall that the hot spots keep changing, but only slightly, with time. So to model the lightcurves, we choose the hot spots at a certain time, and then assume those hot spots to be unchanging.
Fig. \ref{real} shows the variability curves for the realistic spots
in the PW case. The neutron star has
$P=3\mbox{ ms}$. We see that general relativistic effects are the
strongest. Even at such a large rotation speed, the effects of
rotation are relatively small, which justifies neglecting frame dragging effects.

For clarity, Fig. \ref{realgauss} shows the final light curves (solid lines)
after including all effects, for different misalignment and inclination angles. The curves are almost sinusoidal for
small inclination angles when Doppler and time-delay distortions
are relatively weaker, and deviate noticeably from a sinusoidal
shape for larger inclination angles. The pulse fractions depend on
the misalignment and inclination angles and the shape of
the hot spots. In most cases the pulse fractions are seen to be
quite large compared to the observed values of a few percent for
real stars. We see small pulse fractions in our simulations when the misalignment angle or the inclination angle is of the order of a few degrees. The most probable reason then for the small pulse fractions of real pulsars is that they have misalignment angles of the order of a few degrees. Another possible explanation is that scattering of
the hot spot radiation by the surrounding matter reduces the
amplitude of oscillations (Brainerd \& Lamb 1987).

\fign{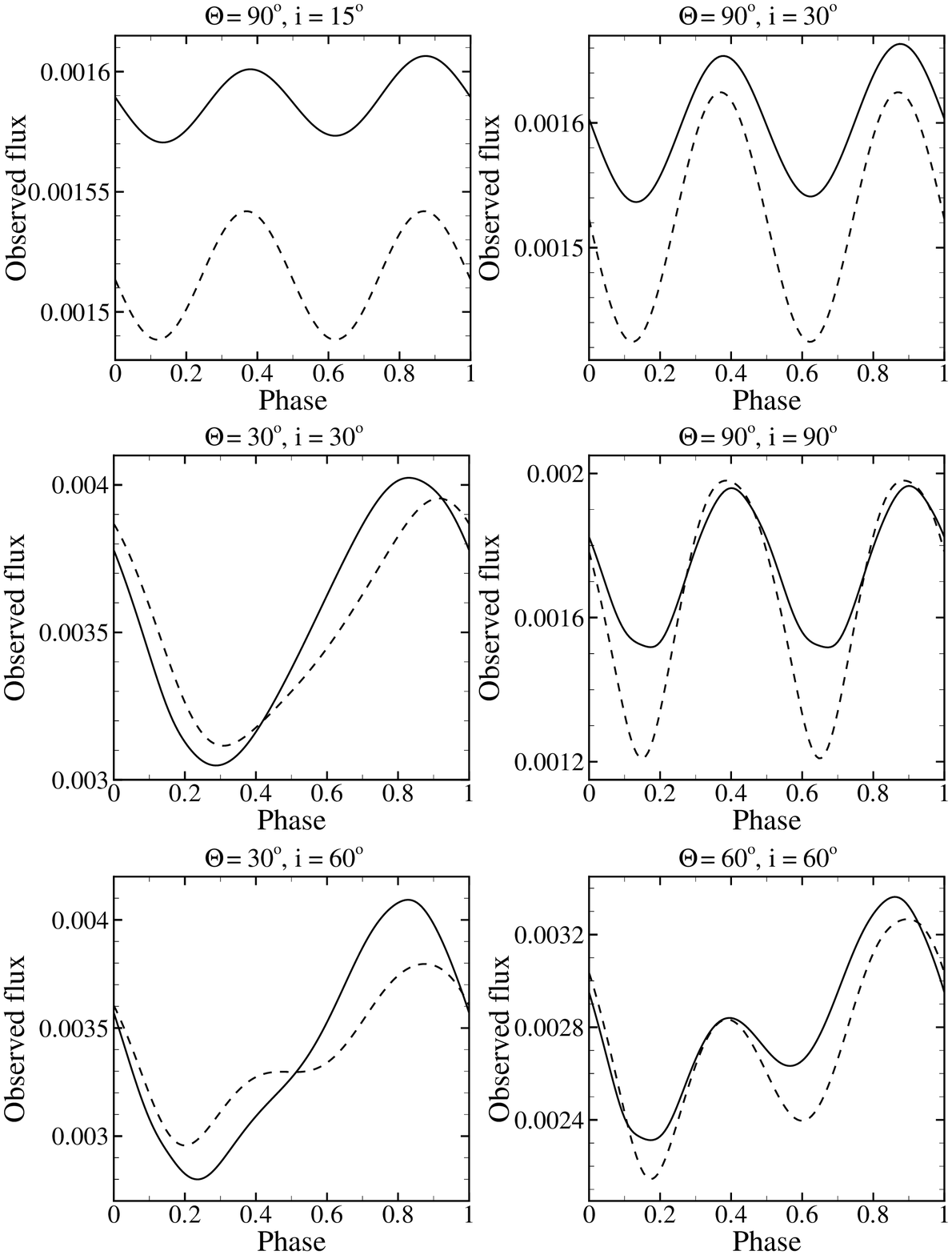}{Final light curves, including all effects (solid lines), with best-fit gaussian-spot lightcurves (dashed lines), for realistic spots in the PW case for 3-ms pulsars. The pulse fractions are 2\% and 8\% respectively for the top two panels, $\sim$ 25-30 \% for the middle two, and $\sim$ 35-40 \% for the bottom two.}{realgauss}

We compared the lightcurves from realistic spots with those from gaussian ones (dashed lines in Fig. \ref{realgauss}). We noted that the hot spots for the $\Theta = 30^\circ$ case are the most amenable to approximation by gaussian spots. So we approximated them with gaussian spots of the same width ($34^\circ$ in this case). The hot spots for the $\Theta = 60^\circ$ and $\Theta = 90^\circ$ cases depart significantly from a round shape, but we tried approximating them with gaussian spots of the same width as for the $\Theta = 30^\circ$ case. The gaussian spots are centered at the magnetic poles in each case. Their intensity in each case was chosen such that the total flux from the star as seen by a distant observer looking at the spot from directly above the magnetic pole would be the same as that in the case of the realistic spots. We also set the two antipodal hot spots to be identical. The lightcurves thus obtained are also shown in Fig. \ref{realgauss}. The fractional r.m.s. errors are $< 5\%$ for small ($\lesssim 60^\circ$) inclination angles and are $\sim$ 5-10 \% for large ($\gtrsim 60^\circ$) inclination angles. The two main reasons for gaussian spots not reproducing the lightcurves perfectly are that the shapes of the realistic hot spots are more complicated than gaussian, and that the two antipodal hot spots are not usually identical.
Fig. \ref{realcomp}a compares the variability curves for two
identical geometries in the PW and non-PW cases. The hot
spots, and hence the variability curves, have similar shapes in
the two cases. We found that the flux in the PW case is a few times higher than that in the non-PW case in general, as expected from the hot spots shown in Fig. \ref{funnelspots}. However, the pulse fractions turned out to be smaller in the PW case.

\fign{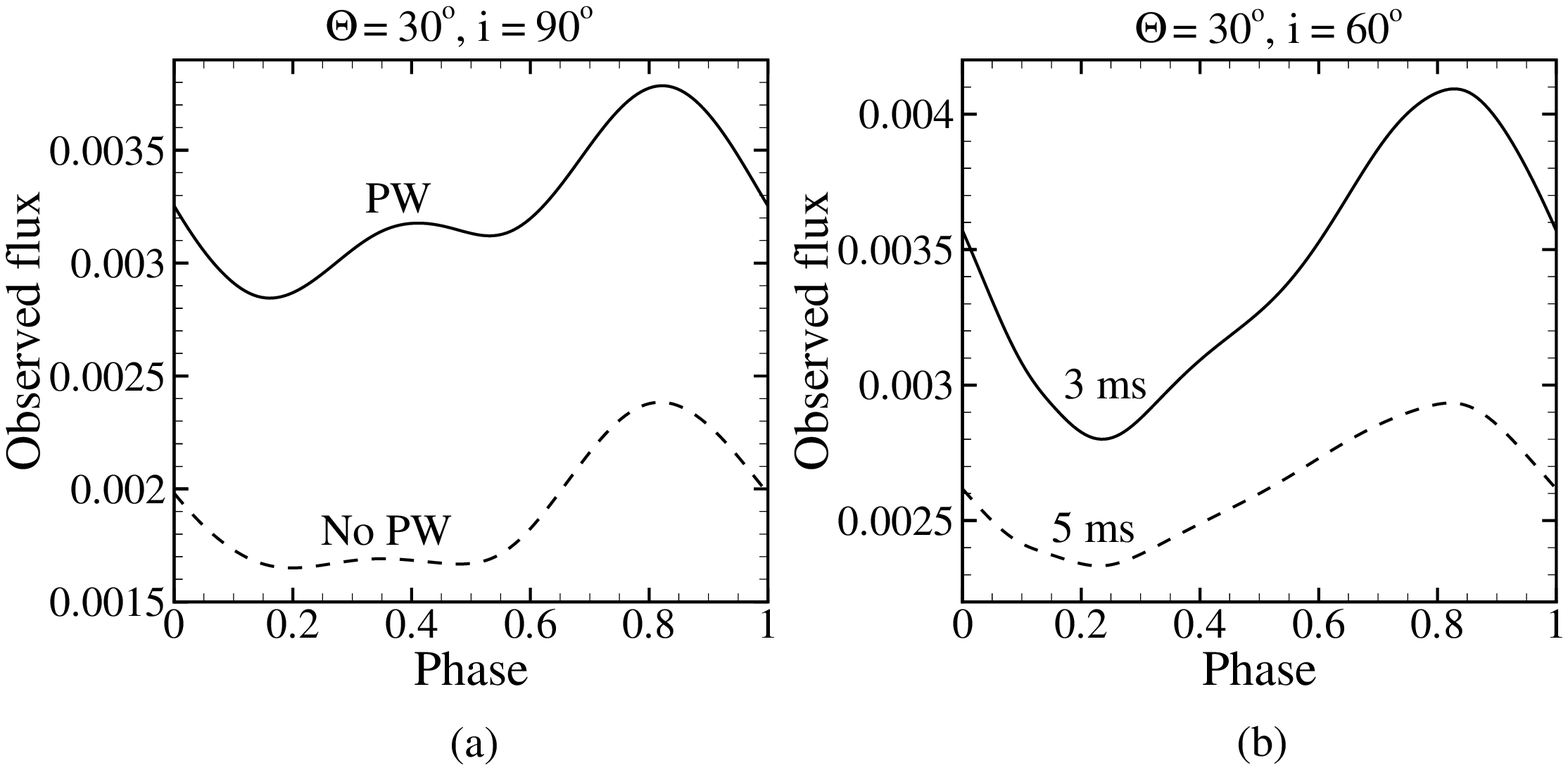}{(a) Comparison of PW (solid line) and non-PW (dashed line) lightcurves, including all effects. (b) Comparison of lightcurves for 3-ms (solid line) and 5-ms (dashed line) pulsars, including all effects.}{realcomp}

\figw{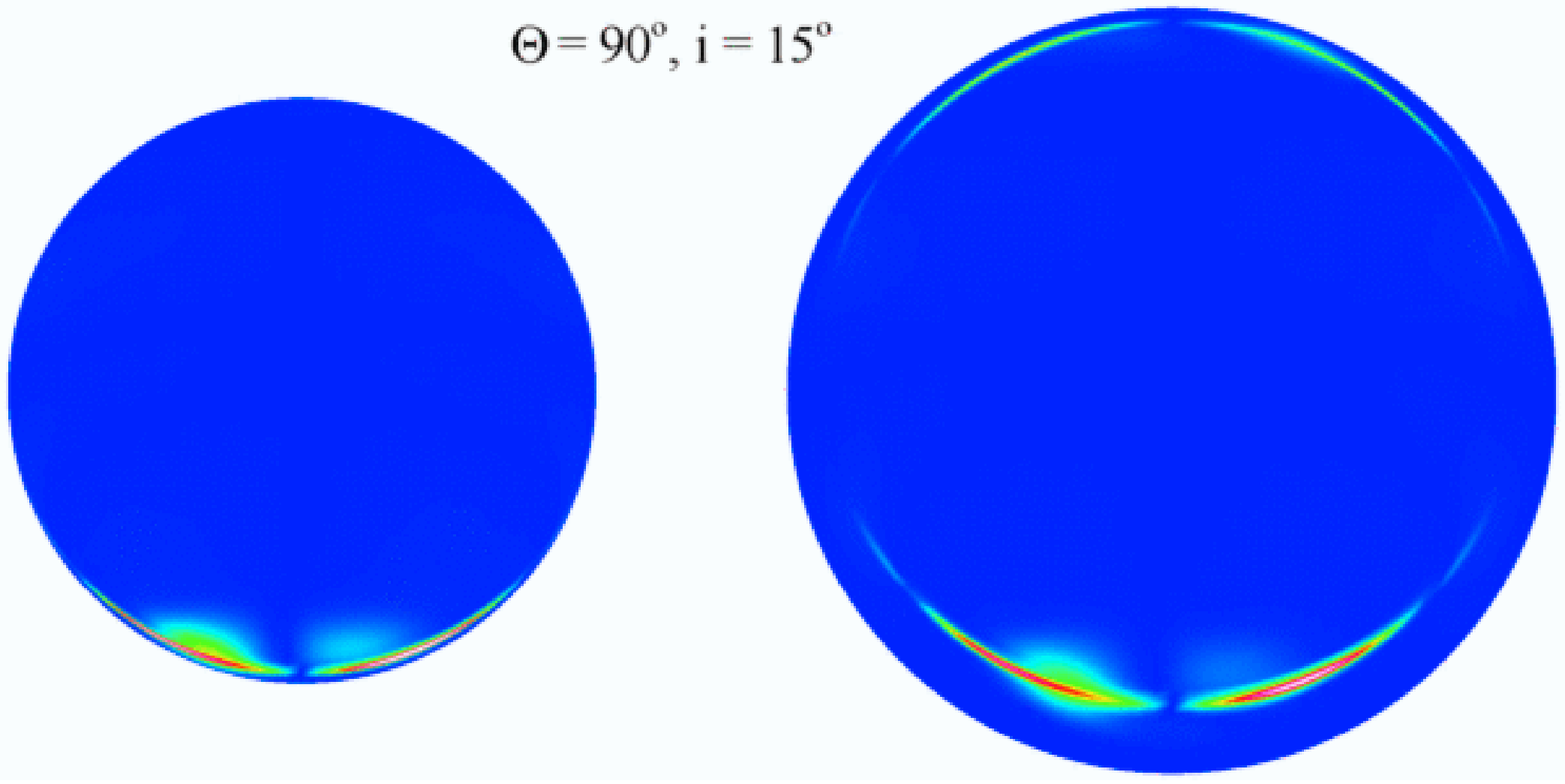}{Observed stellar image, without including any effects (left panel) and including general relativistic effects (right panel). Note the apparent enlargement of the stellar image and the simultaneous visibility of both antipodal hot spots because of general relativistic effects.}{image}

We also compared the variability curves of 3-ms and 5-ms neutron
stars for two identical geometries. The results are shown in Fig.
\ref{realcomp}b. The shapes of the hot spots and curves are again similar in these two cases. Both the average flux and pulse fraction are higher for the 3-ms star.

We plotted the observed stellar image for our fiducial 3-ms pulsar
in the PW case with and without general relativistic effects,
for $\Theta = 90^\circ$ and $i=15^\circ$, shown in Fig.
\ref{image}. We see that both the antipodal hot spots are visible
simultaneously in this case. Fig. \ref{lighttravel}b shows that
due to bending of light, the star should look larger than it
actually is, which is also seen here. We do not take the Doppler
and time delay effects into account in these images, since those
effects are not noticeable in the images.

\section{Conclusions}
We modeled the lightcurves of accreting millisecond pulsars using hot spots obtained in full 3D MHD simulations done using the Paczy\'nski-Wiita potential, for 3 ms and 5 ms-period pulsars. We found that the main effect of the
Paczy\'nski-Wiita potential is to {\it increase the accretion rate and the emitted flux}. The variability curves in our model are
strongly affected by general relativistic effects, and to a lesser
extent by Doppler, time-delay and frame dragging effects. General
relativistic effects decrease the pulse fraction, while Doppler
and light-travel-time effects increase it and distort the light
curve. The amount by which the pulse fraction changes, and the
distortion of the light curve, depend on the hot spots' position and shape, which are determined by the misalignment angle, and on the observer inclination angle. The small pulse fractions observed in real pulsars suggests that they might have small misalignment angles, of the order of a few degrees.

We compared the lightcurves from the realistic hot spots that we obtained from our MHD simulations with those from simple hot spots with a gaussian flux distribution centered at the magnetic poles. We found that the gaussian-spot lightcurves differ from the realistic-spot lightcurves by $< 10\%$, and that therefore gaussian spots are a reasonable approximation for the realistic ones. We plan to investigate in the future how the size and intensity of these equivalent gaussian spots depend on physical parameters pertaining to the star and the disk.

Numerical variability models, along with analytical models, are
important tools for studying periodic and quasi-periodic
oscillations from X-ray pulsars. In the region near the star, hot
spots, funnel streams and features in the accretion disk like
density waves could be responsible for producing these
oscillations. Our 3D simulations are useful for studying these
features. However, a more accurate model of the variability
curves, which takes into account emission and absorption of
radiation by the accreting matter, and temporal changes in the hot
spots, is needed to determine the role of these features in
producing the oscillations. Also, one of the major simplifications
in our model is that we ignore the dynamics of the matter after it
falls on the star's surface. This is not a good assumption for
millisecond pulsars, where thermonuclear burning of the matter
falling on the stellar surface is a possibility (Joss \& Li 1980;
Bildsten \& Brown 1997; Bhattacharyya et al. 2005), which could
change the dynamics of the matter flow and magnetic field around
the star, especially since the magnetic field is relatively weak.
In that case, the hot spots would serve to give an idea of the
place where matter would accumulate and thermonuclear burning
could start.

\acknowledgements This work was supported in part by NASA grants
NAG5-13220, NAG5-13060, and by NSF grant AST-0307817. The authors
thank Drs.\ A. V. Koldoba and G. V. Ustyugova for developing the
code used in our simulations, the referee for valuable comments which greatly improved the paper, and Dr.\ J. Poutanen for valuable
suggestions.

\end{document}